\documentclass[12pt,preprint]{aastex}
\begin{document}

\title{Photometric Orbits of Extrasolar Planets}
\author{Robert A.\ Brown}
\affil{Space Telescope Science Institute, 3700 San Martin Drive, Baltimore, MD 21218}
\email{rbrown@stsci.edu}

\begin{abstract}
We define and analyze the photometric orbit (PhO) of an extrasolar planet observed in reflected light. In our definition, the PhO is a Keplerian entity with six parameters: semimajor axis, eccentricity, mean anomaly at some particular time, argument of periastron, inclination angle, and effective radius, which is the square root of the geometric albedo times the planetary radius. Preliminarily, we assume a Lambertian phase function. We study in detail the case of short-period giant planets (SPGPs) and observational parameters relevant to the \emph{Kepler} mission: 20~ppm photometry with normal errors, 6.5~hour cadence, and three-year duration. We define a relevant ``planetary population of interest'' in terms of probability distributions of the PhO parameters. We perform Monte Carlo experiments to estimate the ability to detect planets and to recover  PhO parameters from light curves. We calibrate the completeness of a periodogram search technique, and find structure caused by degeneracy. We recover full orbital solutions from synthetic \emph{Kepler} data sets and estimate  the median errors in recovered PhO parameters. We treat in depth a case of a Jupiter body-double. For the stated assumptions, we find that \emph{Kepler} should obtain orbital solutions for many of the 100--760 SPGP that Jenkins \& Doyle (2003) estimate \emph{Kepler} will discover. Because most or all of these discoveries will be followed up by ground-based radial-velocity observations, the estimates of inclination angle from the PhO may enable the calculation of true companion masses: \emph{Kepler} photometry may break the ``$m \sin i$'' degeneracy. PhO observations may be difficult. There is uncertainty about how low the albedos of SPGPs actually are, about their phase functions, and about a possible noise floor due to systematic errors from instrumental and stellar sources. Nevertheless, simple detection of SPGPs in reflected light should be robust in the regime of \emph{Kepler} photometry, and estimates of all six orbital parameters may be feasible in at least a subset of cases.
\end{abstract}

\section{Introduction}

\begin{figure}
\epsscale{.8}
\plotone{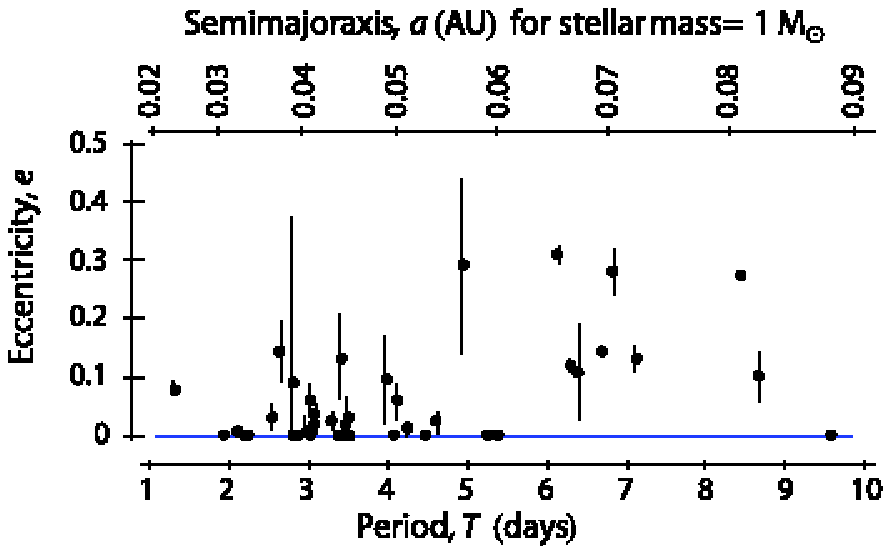}
\caption{Eccentricities and periods of the 40 extrasolar planets with period $T\leq 10$ days, as listed in the catalogue at the 
California \& Carnegie Planet Search webpage (http://exoplanets.org/) on March 23, 2006.}\label{f1}
\end{figure}

We address \emph{Kepler}'s observations of ``photometric orbits'' (PhOs), which is our term for the variable flux of reflected starlight from short-period giant planets (SPGPs). We believe that an improved understanding of this enigmatic class of objects will be an important legacy of the \emph{Kepler} mission. We have conducted a ministudy of \emph{Kepler}'s expected performance in detecting SPGPs and in estimating their orbital parameters via the PhO. We have found unexpected scientific benefits. 

A ``Keplerian data set'' is an observational record governed by the classical motion of two bodies interacting by the force of gravity. Radial-velocity, astrometric, and photometric observations of extrasolar planets are prime examples. In fact, the three types provide overlapping and complementary information, which creates synergism when more than one data type is available, as we can expect in many cases for SPGPs discovered by \emph{Kepler}---at least for radial velocity and photometry (including both transits and reflected light).

When first discovered in the 1990s, SPGPs were a surprise, and we still do not understand their basic properties. Why do they exist in such abundance? How were they created? After creation, did they migrate in distance from the star, and if so, by what mechanism? Why do the orbits of many SPGPs have significant eccentricity, if tides should efficiently circularize them? (See Fig.~\ref{f1}.) What are the optical radii and actual masses of SPGPs, and what are their physical conditions? Of what types of stellar and planetary systems are SPGPs members?

\begin{figure}
\plotone{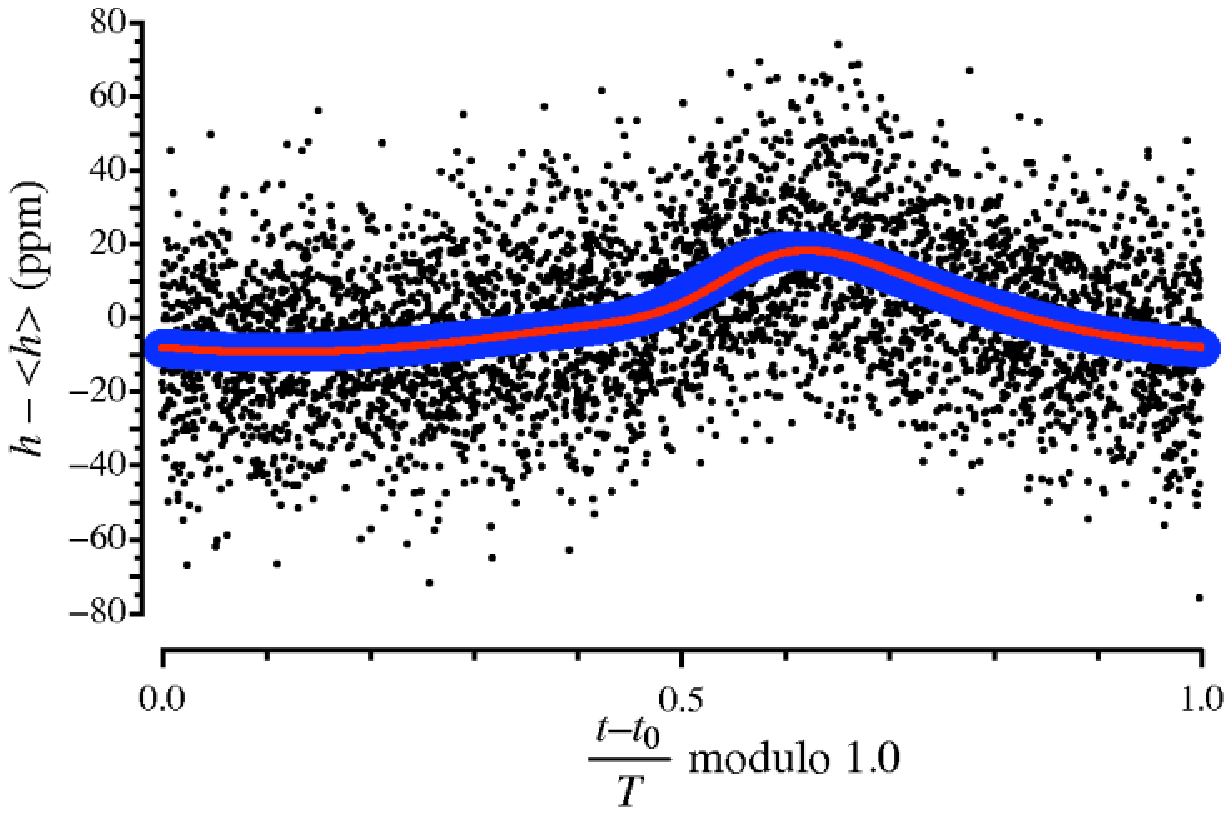}
\caption{Simulated data set created for the Jupiter body-twin with 20~ppm noise. The mean has been subtracted, and the data is folded on the orbital period. The curves show the true (red) and best-estimate (blue) solutions. We treat this data set in depth in \S8. Although one's first impression might be that this is a very noisy data set, in fact the signal-to-noise ratio is $S/N=48$. ($S/N$ is discussed in \S10.) The planet is easily detected by \emph{Kepler} with the assumed performance, and all orbital parameters are robustly estimated (although there are systematic effects due to uncertainty in the phase function; see \S9). } \label{old-f10}
\end{figure}

In the next decade, new information from a variety of sources will help address these questions, and we expect that \emph{Kepler}'s observations will be important. \emph{Kepler}'s discoveries will increase the number of known SPGPs, and its measurements of PhOs will increase the value of other data sets by independent estimates of some planetary parameters, and will provide some information not otherwise available.

The current \emph{Kepler} science team plans to search for SPGPs by seeking peaks in the periodograms (power spectra) of stellar light curves  and to determine geometric albedos for transiting SPGPs (Jenkins \& Doyle 2004). This paper goes well beyond those modest ambitions.

We analyzed many simulated \emph{Kepler} data sets to explore how reflected-light observations might provide additional information about SPGPs.  (Fig.~\ref{old-f10} shows one example, treated in depth in \S8.) These Monte Carlo experiments allow us (1)~to provide a preliminary calibration of the completeness of the periodogram search technique. Such calibration is generally informative and will enable estimates of the true occurrence frequency of SPGPs as a function of planetary parameters. A sufficiently strong signal allows us (2)~to provide \textit{complete solutions} of the PhO. So far as we can detect, no one else has previously thought this possible. 

The benefits of full PhO solutions include: (1)~using the estimated orbital inclination to determine the true planetary mass when radial-velocity measurements are available; (2)~improved albedo estimates for eccentric transiting planets, by allowing data points to be interpreted with their correct planetary phase angles; (3) independent estimates of the four orbital parameters in common with radial velocity; and (4)~possible estimates of the planetary phase function in very high signal-to-noise cases. 

PhO observations may be difficult. There is uncertainty about how low the 
albedos of SPGPs actually are, about their phase functions, and about a possible 
noise floor due to systematic errors from instrumental and stellar sources. 
Nevertheless, simple detection of SPGPs in reflected light should be robust in 
the regime of \emph{Kepler} photometry, and estimates of all six orbital parameters may be feasible in at least a subset of cases.


\section{Theory of the PhO}

To the best of our knowledge, a theory of the PhO does not exist in the literature, and so we provide a summary account.

The most convenient observable of the PhO is based on---but not exactly equal to, as we will see---the ratio of planetary to
stellar flux, which is the function:
\begin{equation}
h \equiv  \frac{R_\mathrm{eff}^2}{r^2}\, \Phi [\beta]~~, 
\end{equation}
where $r$ is the radial distance between the star and planet; $R_\mathrm{eff}$ is the effective planetary radius
\begin{equation}
R_\mathrm{eff} \equiv \sqrt{p}\, R_p~~,
\end{equation}
where $p$ is the geometric albedo, and $R_\mathrm{p}$ is the planetary radius; and $\Phi[\beta]$ is the phase function, where $\beta$ is the planetary phase angle, which is the planetocentric angle between the star and observer. Nominal values of $p$ are 0.66 for a conservative Lambert sphere (a ping-pong ball) and 0.5 for Jupiter at visible wavelengths. Thus, for Jupiter, $R_\mathrm{eff} = 
7.9~R_{\oplus }$, in terms of the Earth radius. 

\begin{deluxetable}{lcccc}
\tablewidth{0pt}
\tablecaption{The ministudy's ``planets of interest''}
\tablehead{
\colhead{Parameter} &\colhead{Minimum} &\colhead{Maximum} &\tablehead{Units} &\colhead{Distribution}}
\startdata
Semimajor axis ($a$) &0.0\rlap{3} &0.0\rlap{9} &AU &uniform\\
Eccentricity ($e$) &0.0 &0.5 &  &uniform\\
Initial mean anomaly ($M_0$) &0\phn\phd &$2\pi$ &radians &uniform\\
Argument of periastron ($\omega$) &0\phn\phd &2$\pi$ &radians &uniform\\
Inclination angle ($i$) &0\phn\phd &$\frac{\pi}{2}$ &radians &uniform on sphere\\
Effective planetary radius ($R_\mathrm{eff}$) &0\phn\phd &\llap{2}0\phn\phd &$R_{\oplus}$ &uniform\\
Period ($T$) &1.9 &9.9 &days &computed from $a$\\
\enddata
\tablecomments{The random deviate for $i$ is $\cos^{-1}$ (1--$\mathcal{R}$), where $\mathcal{R}$ produces pseudorandom numbers in the range 0--1.}
\end{deluxetable}

We expect $\Phi[\beta]$ to decrease monotonically from 1 to 0 over the range of $\beta$ from 0 to $\pi$. In the absence of better information, we adopt the Lambertian phase function in the ministudy:
\begin{equation}
\Phi _\mathrm{L}[\beta] = \frac{\sin  \beta  + (\pi - \beta ) \cos  \beta}{\pi }~~.
\end{equation}
With superior data, the opportunity may arise to introduce into the analysis, and to estimate, one or more additional PhO parameters associated with the phase function. In \S10 we discuss the systematic errors introduced when a data set is analyzed with the wrong phase function.

As usual, the planetary position relative to the star is fully described by seven parameters: $a~(0-\infty)$, the semimajor axis of the orbit with a focus at the star; $e~(0-1)$, the orbital eccentricity; $M_0~(0-2\pi)$, the mean anomaly at an arbitrary $t_1$, taken here to be the time of the first observation; $\omega~(0-2\pi)$, the argument of periastron; $i~(0-\pi)$, the inclination angle; $\Omega~(0-2\pi)$, the position angle of the ascending node; and $T~(0-\infty $), the orbital period.

\emph{Kepler}'s photometric observations are spatially unresolved and therefore do not constrain $\Omega$, which we set to zero without loss of generality. Similarly, $i$ is nondegenerate only in the range $0-\pi/2$.

If we can estimate the sum of the planetary and stellar masses, $m_\mathrm{p}$ and $m_\mathrm{s}$, then Kepler's Third Law, informed by Newton's Law of Gravity, provides a relationship between $a$ and $T$:
\begin{equation}
a^3= \frac{G\left(m_\mathrm{p}+m_\mathrm{s}\right)}{4 \pi ^2} T^2~~,
\end{equation}
where $G$ is the gravitational constant. We take advantage of Eq.~(4) to retire $T$ as an independent parameter by assuming that we
can neglect $m_\mathrm{p}$ relative to $m_\mathrm{s}$, and that we will have an estimate of $m_\mathrm{s}$ available from the spectrophotometric characteristics of the star. 

To compute the planetary position in space, which is needed to compute $r$ and $\beta$ in Eq.~(1), we establish a Cartesian
coordinate system $\{x,\,y,\,z\}$ with its origin on the star. The plane of the sky is the domain $\{x,\,y,\,0\}$, with +$x$ north, +$y$ east, and +$z$ in the direction of the observer. The base $\{a,\,e\}$ orbit, with $\omega = 0$ and $i = 0$, lies in the plane of the sky, with periastron located on the +$x$ axis, and the motion of the planet at periastron is in the +$y$ or east direction. The base orbit is direct, meaning that position angle always increases with time. The first Eulerian angle, $\omega$, is a right-hand rotation around +$z$ axis, which moves periastron to the east. The second Eulerian angle, $i$, is a left-hand rotation around the +$x$ axis.

The planetary position in the base orbit at time $t$ is
\begin{equation}
\{x_\mathrm{b},\,y_\mathrm{b},\,z_\mathrm{b}\} = \{r\,\cos\nu,\ r\,\sin\nu,\ 0\}~~, 
\end{equation}
where the radial distance between the star and planet is
\begin{equation}
r = \frac{a \left(1 - e^2\right)}{(1 + e\,\cos\nu)}~~,
\end{equation}
where $\nu$ is the true anomaly, which is the root of the equation
\begin{equation}
\tan \frac{\nu }{2} = \sqrt{\frac{(1 + e)}{(1-e)}}\, \tan \frac{E}{2}~~,
\end{equation}
where $E$ is the eccentric anomaly, the root of Kepler's Equation:
\begin{equation}
E - e \sin\,E = M~~,
\end{equation}
and $M$ is the mean anomaly:
\begin{equation}
M = M_0 + 2\pi \frac{t-t_1}{T}~~,
\end{equation}
where $M_0$ is the initial mean anomaly, and $t_1$ is the time of the first observation.

The planet's position in space is then
\begin{eqnarray}
\{x,\,y,\,z\}&=&r\{-\sin\nu\,\sin\omega + \cos\nu\,\cos\omega,\,\cos\nu\,\cos i\,\sin\omega + \sin\nu\,\cos i\,\cos\omega,\nonumber\\
&&\quad-\cos\omega\,\sin i\,\sin\nu - \cos\nu\,\sin i\,\sin\omega\}~~,
\end{eqnarray}
from which we can compute:
\begin{equation}
\beta = \arctan \frac{-\sqrt{x^2+ y^2}}{z}~~.
\end{equation}
The results $r$ and $\beta$ from Eqs.~(6) and (11) permit the computation of $h[t]$ in Eq.~(1).

\S8 provides a particular, in-depth example of a PhO.

\section{Keplerian Data Sets of the PhO}

In the ministudy, we constructed each simulated data set from particular values for ten parameters, which served as the controls of our Monte Carlo experimentation. The parameters, grouped in three subsets, were generated as follows:  
(1)~PhO parameters $\{a,\,e,\,M_0,\,\omega,\,i,$ $R_\mathrm{eff}\}$ were drawn from appropriate random deviates over ranges relevant to \emph{Kepler}'s SPGPs (see Table~1). These ranges and deviates define ``the planets of interest.''  (2)~We chose the three observational parameters, $\{\sigma,\,\textit{duration},\,\textit{cadence}\}=\{2\times10^{-5}$, 3~years, 6.5~hours$\}$, where we used the values of photometric uncertainty, $\sigma$, and \textit{cadence} provided in NASA Research Announcement NNH07ZDA001N--KPS, soliciting Kepler Mission Participating Scientists. This document states:  ``The mission will achieve a photometric precision of 20 parts-per-million (ppm) on a $m_\mathrm{v}=12$ magnitude G2V star, for a 6.5-hour  integration and using differential ensemble processing.'' Our choice of the duration of the observations is arbitrarily 75\% of the planned mission duration.  (3)~With no particular star in mind, we chose $m_\mathrm{s} = 1~M_{\odot}$, in terms of the solar mass.

Each simulated data set had the form
\begin{equation}
\left\{t,\,\hat{h}\right\} \equiv \left\{\left\{t_1,\,\hat{h}_1\right\},\,\left\{t_2,\,\hat{h}_2\right\}, ... \left\{t_N,\,\hat{h}_N\right\}\right\}~~,
\end{equation}
containing a number of data points $N$ equal to the integer part of $\frac{\textit{duration}}{\textit{cadence}}$, or $N=4046$.  The values of $\hat{h}$ were generated by a normal random deviate with mean value equal to the value of $h$ in Eq.~(1) at the value of $t$ and with standard deviation $\sigma$. In other words, the value of $h$ was not smoothed  over a 6.5~hour interval.

\section{Periodogram Planet Search}

At least for non-transiting SPGPs, \emph{Kepler}'s first indication of a planet will come from a periodogram analysis. Also, the information in the periodogram about the planetary period (semimajor axis) will provide the starting point for finding a full solution for the PhO. For these reasons, understanding the strengths and limitations of the periodogram is a cornerstone of PhO research. In addition, we recognize an opportunity to calibrate the periodogram in terms of search completeness, which will enable the estimation of the underlying SPGP population.

The periodogram is the Fourier transform of the autocorrelation function. In our normalization, the estimated periodogram \textit{pg} is the vector
\begin{equation}
pg[\textit{freq}_j] \equiv \frac{1}{2\pi } \left|\hat{G}\left(\textit{freq}_j\right)\right|^2~~, 
\end{equation}
where $\hat{G}$ is the discrete Fourier transform of the data record:
\begin{equation}
\hat{G}[\textit{freq}_j] \equiv \frac{1}{\sqrt{N}} \sum _{k=1}^N \hat{h}_k\,e^{2\pi\,i\,\textit{freq}_j\,t_k}~~,
\end{equation}
where
\begin{equation}
\textit{freq}_j \equiv \frac{j}{N\,\textit{cadence}}~~,
\end{equation}
and $j = 0, 1, 2 ... \frac{N}{2}$ indexes the non-redundant frequencies. The highest value, \textit{freq}$_{N/2} = 
\frac{1}{2\,\textit{cadence}}$, is the Nyquist frequency of critical sampling.

The ministudy's protocol was, first, to smooth the periodogram with a Daniell window of second order; second, to locate the highest point in the periodogram for $j>0$ and record its frequency (\textit{pgFreq}) and peak value (\textit{pgPeak}); and third, to determine if the planet was ``detected'' by this test:
\begin{equation}
\textit{pgFreq} \stackrel{?}{=} \frac{1}{T},\, \frac{2}{T},~\mathrm{or}~\frac{3}{T}~~,
\end{equation}
within 2\%. (For higher $e$, dominant spectral power may be transferred from the fundamental frequency to harmonics.)
 
\begin{figure}
\plotone{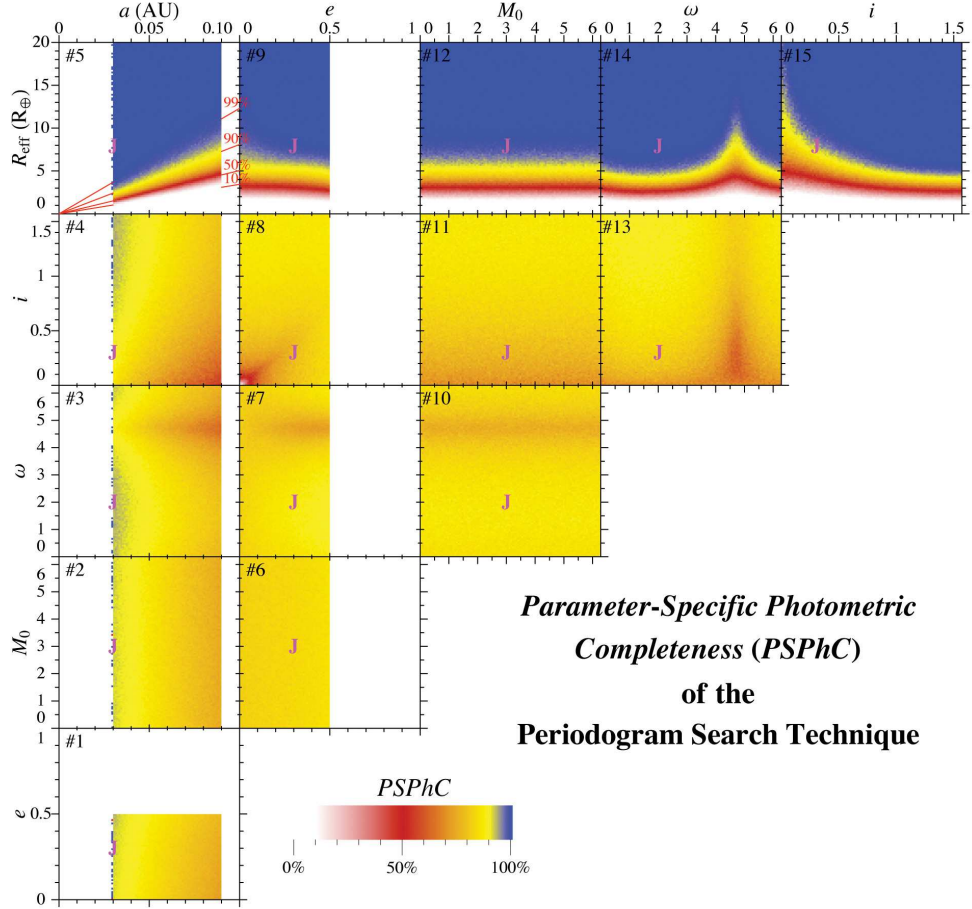}
\caption{Ministudy results for the \textit{parameter-specific photometric completeness} (\textit{PSPhC}) of periodogram searches for extrasolar planets. As shown by the red lines in frame~5, \textit{PSPhC}~= 10, 50, 90, and 99\% is achieved for $R_\mathrm{eff}/a\approx 35$, 53, 91 \& 161 $R_{\oplus }$/AU, respectively. This functional dependence is expected from Eqs.~(1) and (6), which indicate the PhO is invariant under changes in $R_\mathrm{eff}$ and $a$ if the value of $R_\mathrm{eff}/a$ is preserved. The magenta ``J'' indicates the Jupiter body-twin studied in \S8.}\label{old-f2}
\end{figure}

Figure~\ref{old-f2} shows the \textit{parameter-specific photometric completeness} (\textit{PSPhC}) of the periodogram search technique. Here,
we created 15 frames of $100\times100$ cells (1\% resolution) to compile and display the results of the Monte Carlo experiment. This ``paramagram'' is an exhaustive two-dimensional projection of parameter space, the frames comprising all possible pairs of PhO parameters. After determining which cell in each frame corresponded to the data set's control values of 
$\{a,\,e,\,M_0,\,\omega,\,i,\,R_\mathrm{eff}\}$, we incremented the ``number tested'' (\textit{nTested}) in that cell by one, and if the planet was detected, we also incremented the ``number detected'' (\textit{nDetected}) in that cell by one. After accumulating all results, we estimated \textit{PSPhC} in each cell using the general definition:
\begin{equation}
\textit{estimated completeness} \equiv \frac{\textit{nDetected}}{\textit{nTested}}~~,
\end{equation}
which is the best estimate of the probability of detecting a planet with parameters in a cell's range (Table~1), assuming its presence.

\textit{PSPhC} is independent of $M_0$, because we are observing a large number of cycles. Therefore, any direct variation---or
lack of variation---of \textit{PSPhC} with respect to another parameter is seen best in the frame of the paramagram it shares with 
$M_0$. For example, frame~6 shows that the periodogram technique is neutral with respect to $e$ over the range of $e$ in Table~1.

The planetary signal vanishes for face-on, circular orbits $\{e=0$, $i=0\}$, which is indicated by the point of low \textit{PSPhC} in the bottom-left corner of frame~8. The streak of degraded \textit{PSPhC} extending diagonally up and right from this corner is due to the ability of increasing $i$, when $\omega\approx\frac{3\pi }{2}$, to cancel increasing signal near periastron from the reduced periastron distance $a(1-e)$ due to $e$ increasing. (Increasing $i$ tilts the dark side of the planet toward the observer most efficiently at $\omega \approx \frac{3\pi }{2}$, as documented by the horizontal streaks of degraded \textit{PSPhC} in frames~3, 7, and 10, and the vertical streaks in frames 13, 14, and 15.) \textit{PSPhC} is reduced because the signals of some of the planets touched by the effect, which were otherwise borderline-detectable, now fail to exceed the detection threshold.

We expect the peak-to-trough amplitude of the PhO, \textit{peakToTrough} $\equiv  h_\mathrm{max} - h_\mathrm{min}$, to correlate
with (1)~detectability via periodogram and (2)~confidence in an estimated orbital solution. In other words, \textit{peakToTrough}---a transcendental function of $\{a,\,e,\,M_0,\,\omega,\,i,\,R_\mathrm{eff}\}$---is also a useful control parameter for our Monte Carlo experiments. That is, \textit{peakToTrough} can be calculated when a synthetic data set is created and carried along to help interpret the results. From Eq.~(1), the maximum value is \textit{peakToTrough} $=8\times10^{-4}$ for $e=0$ for the planetary population defined by Table~1.

\begin{figure}
\plotone{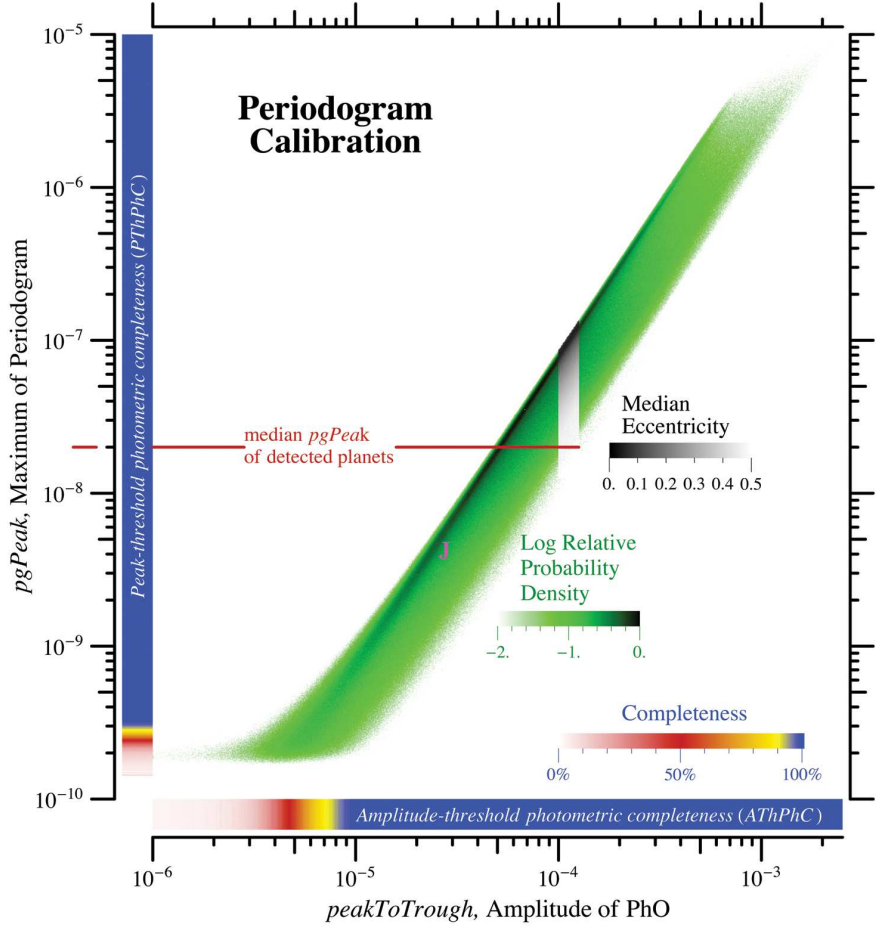}
\caption{In green, the distribution of log-relative probability density for detected planets on the plane spanned by 
\textit{peakToTrough} (``amplitude'') and \textit{pgPeak} (``peak''). The color-coded bands along the ordinate and abscissa show the completenesses \textit{PThPhC} and \textit{AThPhC}. The grayscale stripe illustrates periodogram-power leakage due to eccentricity. The red ocher line indicates the median value of \textit{pgPeak} for detected planets, $2\times10^{-8}$. In \S7, we discuss the universe of full orbital solutions for the population of detected planets above this line. In \S8, we discuss the details of one particular planet, the Jupiter body-twin indicated by the magenta ``J''.}\label{old-f3}
\end{figure}

Figure~\ref{old-f3} plots the $10^7$ periodogram results versus \textit{peakToTrough} and \textit{pgPeak}. In green, we show the log probability
density of \textit{detected} planets parsed on a $1000\times1000$ grid. In the sinusoidal approximation of $h$--$h_\mathrm{min}$, the second-order power law evidenced in Figure~\ref{old-f3}, \textit{pgPeak} $\propto$ \textit{peakToTrough}$^2$, is expected from Parseval's
Theorem:
\begin{equation}
\sum _{j=0}^{N/2} \textit{pg}\left[\textit{freq}_j\right] = 
\frac{1}{4\pi} \sum _{k=1}^N \left|\hat{h}_k\right|^2 \approx \frac{N}{32 \pi}\,\textit{peakToTrough}^2~~,
\end{equation}

\noindent because
\begin{equation}
\sum_{j=0}^{N/2} \textit{pg}\left[\textit{freq}_j\right] \propto \textit{pgPeak}~~,
\end{equation}
where the constant of proportionality depends only on the chosen smoothing of the periodogram. The minimum detectable value of \textit{peakToTrough} is seen to be approximately 2$\sigma$.

Along a vertical at any value of \textit{peakToTrough} in Figure~\ref{old-f3}, the cases with higher eccentricities are smeared downward from the upper-left edge of the probability density distribution, due to increasing leakage of power from the fundamental frequency into harmonics as $e$ increases.

For Figure~\ref{old-f3}, search completeness was again computed via Eq.~(17), but this time specifying \textit{peakToTrough} and \textit{pgPeak}, without regard to the PhO parameters. We call these quantities the \textit{amplitude-threshold photometric completeness} (\textit{AThPhC}) and \textit{peak-threshold photometric completeness} (\textit{PThPhC}), shown by the color-coded bands along the abscissa and ordinate. For example, we find \textit{AThPhC}~$= 0.5$ for \textit{peakToTrough}~$= 4.5\times10^{-6}$ and 
\textit{PThPhC}~$= 0.5$ for \textit{pgPeak}~$= 2.5\times10^{-10}$. (The second number depends on the choice of smoothing, but the first number is universal, corresponding to $S/N=7$; see discussion of $S/N$ is \S10.)

The fraction of all $10^7$ planets in the ministudy that were detected according to Eq.~(16) is 0.91. This quantity is the estimated
\textit{ensemble photometric completeness} (\textit{EPhC}), where the ``ensemble'' is the planetary population of interest defined in Table~1.

The benefits of studying the search completeness of extrasolar planetary observations are fourfold, at least (Brown 2004a, 2005, 2009). First, such study establishes realistic expectations by clarifying search power in objective terms. Second, it provides a scientific metric to inform trades during mission development, and to optimize variable aspects of the observations in the operational period. Third, completeness permits the underlying planetary population to be estimated, by taking into account intrinsically equivalent planets that happen to be undetectable because of unfavorable extrinsic characteristics. (We learn where planets can hide, how many are hiding, and compensate statistically.) Fourth, completeness studies offer unique insights into the complexities of Keplerian observations, as well as glimpses of their intellectual beauty---and occasional obscurities.

\section{The Observable \emph{h}--\boldmath{$<$}\emph{h}\boldmath{$>$}}

Rather than $h$, \emph{Kepler} will actually measure the \textit{sum} of the planetary and stellar fluxes, which means subtracting an
estimate of the stellar flux from the observation, and then dividing, in order to approach $h$.

To complicate matters, $h$ has a zero-point issue, because the minimum of the planetary flux---the pedestal of signal on which the variable part of the planetary light curve stands---is indistinguishable from steady starlight. Currently, we think the proper way to address this issue is to use the \textit{zero-mean} planetary signal, $h$--$<\!\!h\!\!>$, when comparing theory and observation in the process of seeking solutions of the PhO.

In the ministudy, we assumed that $\sigma=20$~ppm applies to measurements of $h$--$<\!\!h\!\!>$, and that the photometric errors are normally distributed.

\section{The Merit Function, \emph{chiSqR}}

For a data set $\{t,\,\hat{h}\}$, we measure the merit of a candidate solution $\{a,\,e,\,M_0,\,\omega,\,i,\,R_\mathrm{eff}\}$ by the sum of the squared, normalized deviations---the ``reduced chi square.'' \textit{chiSqR} is a numerical function with two sets of arguments, the solution and the observational data:
\begin{equation}
\textit{chiSqR}\left[\left\{a,\,e,\,M_0,\,\omega,\,i,\,R_\mathrm{eff}\right\}, \left\{t,\,\hat{h}\right\}\right] \equiv 
\sum_{k=1}^N \frac{1}{N-6} \left(\frac{\left(h\left[t_k\right] - <\!\!h\!\!>\right) - \left(\hat{h}_{k}\, - <\!\!\hat{h}\!\!>\right)}{\sigma}\right)^2~~
\end{equation}

If a global minimum of \textit{chiSqR} can be discovered, its coordinates are the ``solution''---the best estimate of the PhO parameters. More generally, surfaces of constant merit in parameter space divide more probable solutions from less probable, which is the principle
that defines confidence regions. If the noise in a Keplerian data set is understood, it may be possible to associate absolute levels of probability with such surfaces and regions. In particular, if the measurement errors are normally distributed, and if we assume the Keplerian theory is correctly applied, then the probability that the true solution is contained within a surface of constant merit can be computed or looked up in a table: it is the cumulative probability of the theoretical $\chi^2$ statistic.

\section{PhO Fits}

To explore the scientific utility of estimating the six PhO parameters from \emph{Kepler} data sets, we investigated the expected errors in the fitted parameters $\{\hat{a},\,\hat{e},\,\hat{M}_0,\,\hat{\omega },\,\hat{i},\,\hat{R}_\mathrm{eff}\}$. We used the Levenberg-Marquardt
algorithm to locate minima of \textit{chiSqR} for a subset of the Monte Carlo planets in the ministudy. The selection criterion for this subset was \textit{pgPeak}~$>2.0\times10^{-8}$, which is true for 50\% of detected planets and 45\% of all planets specified
by Table~1. 

To save time searching for propitious starting points, the initial estimates of the PhO parameters were the true solutions. 

\begin{figure}
\plotone{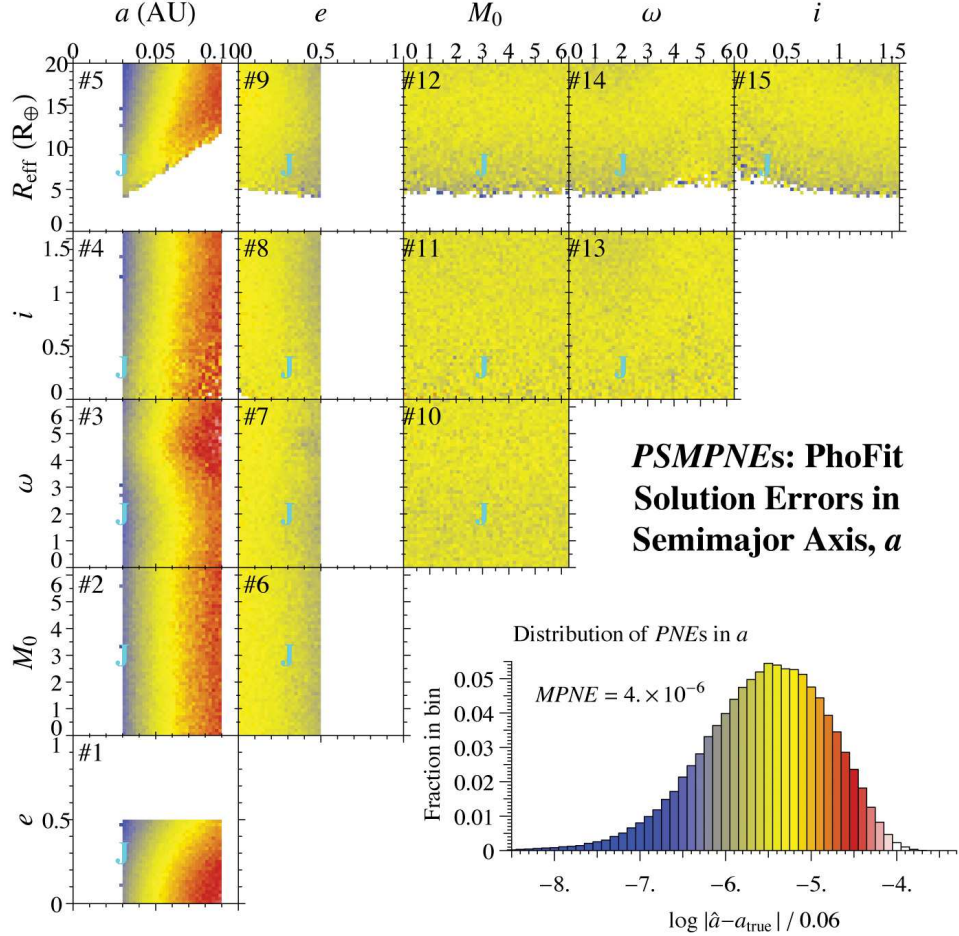}
\caption{The \textit{parameter-specific median projected normalized errors} (\textit{PSMPNEs}) for $a$. Any cell in any grid shows the median error in the subject parameter for the subset of a simulated data sets with \textit{pgPeak}~$>1.5\times10^{-8}$ and with true parameters falling in that cell.The color key is provided by the histogram,which reports the distribution of projected normalized errors 
(\textit{PNEs}) by log value. The cyan ``J" indicates the body-twin of Jupiter studied in \S8.}\label{old-f4}
\end{figure}

\begin{figure}
\plotone{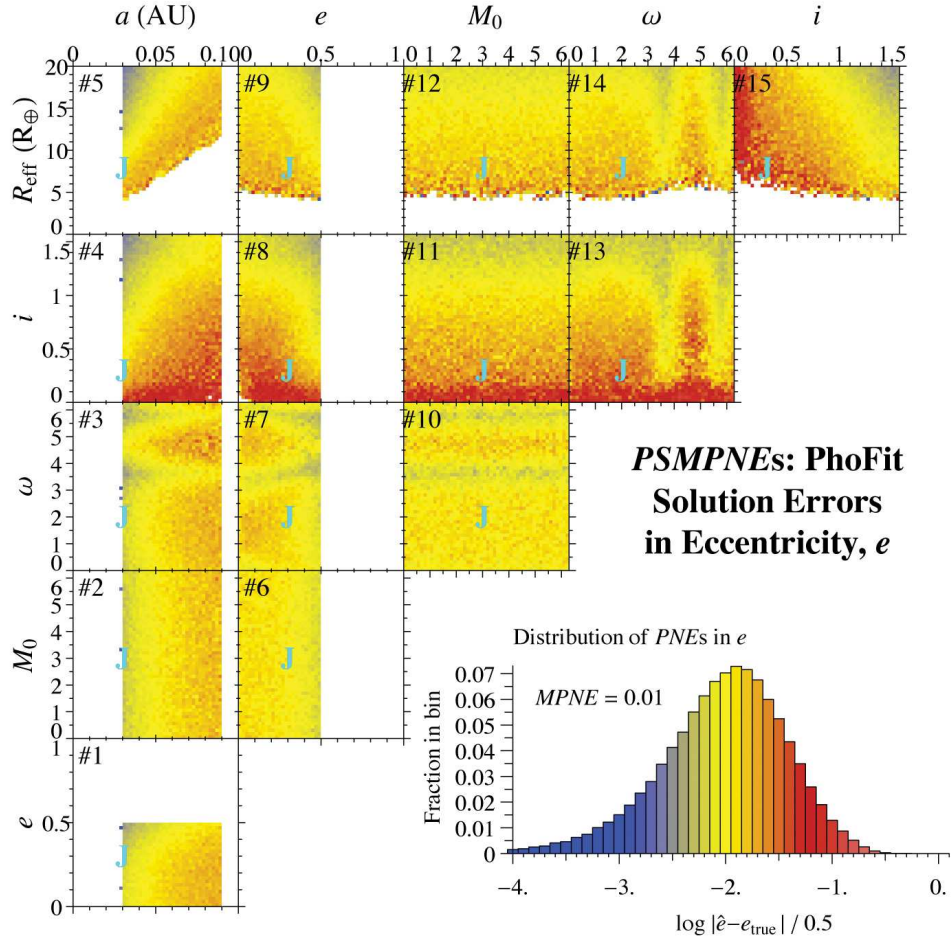}
\caption{\textit{PSMPNEs} for $e$.}\label{old-f5}
\end{figure}

\begin{figure}
\plotone{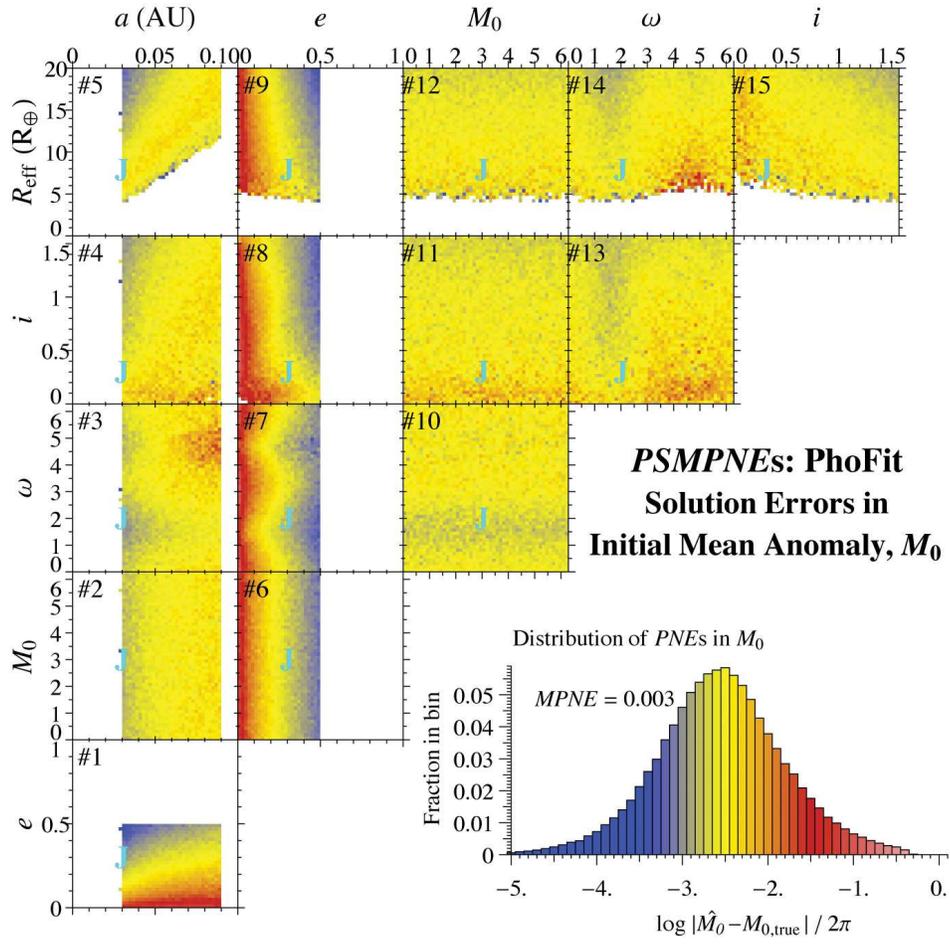}
\caption{\textit{PSMPNEs} for $M_0$.}\label{old-f6}
\end{figure}

\begin{figure}
\plotone{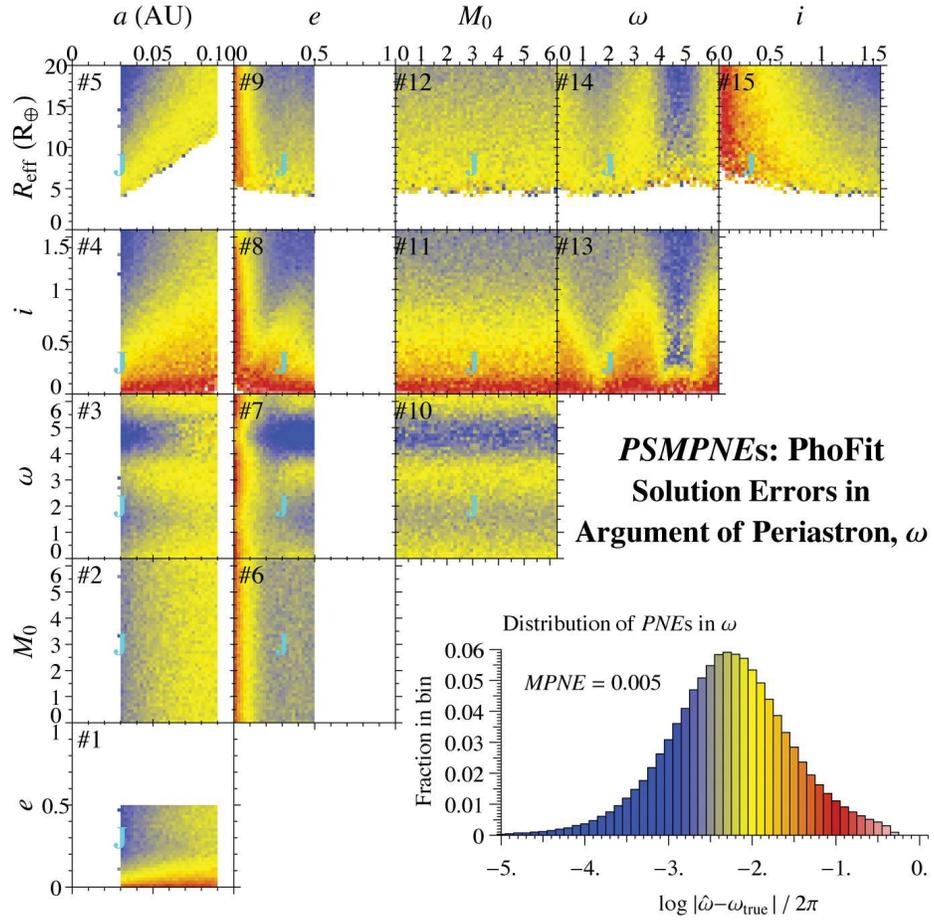}
\caption{\textit{PSMPNEs} for $\omega$.}\label{old-f7}
\end{figure}

\begin{figure}
\plotone{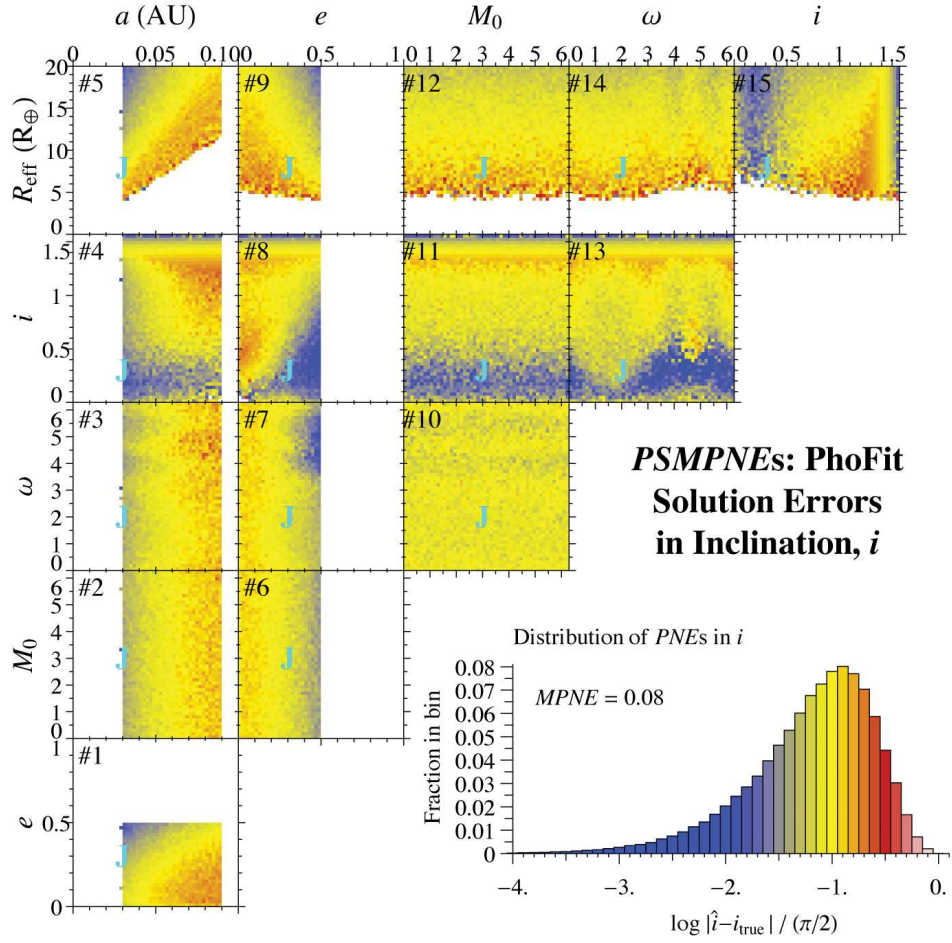}
\caption{\textit{PSMPNEs} for $i$.}\label{old-f8}
\end{figure}

\begin{figure}
\plotone{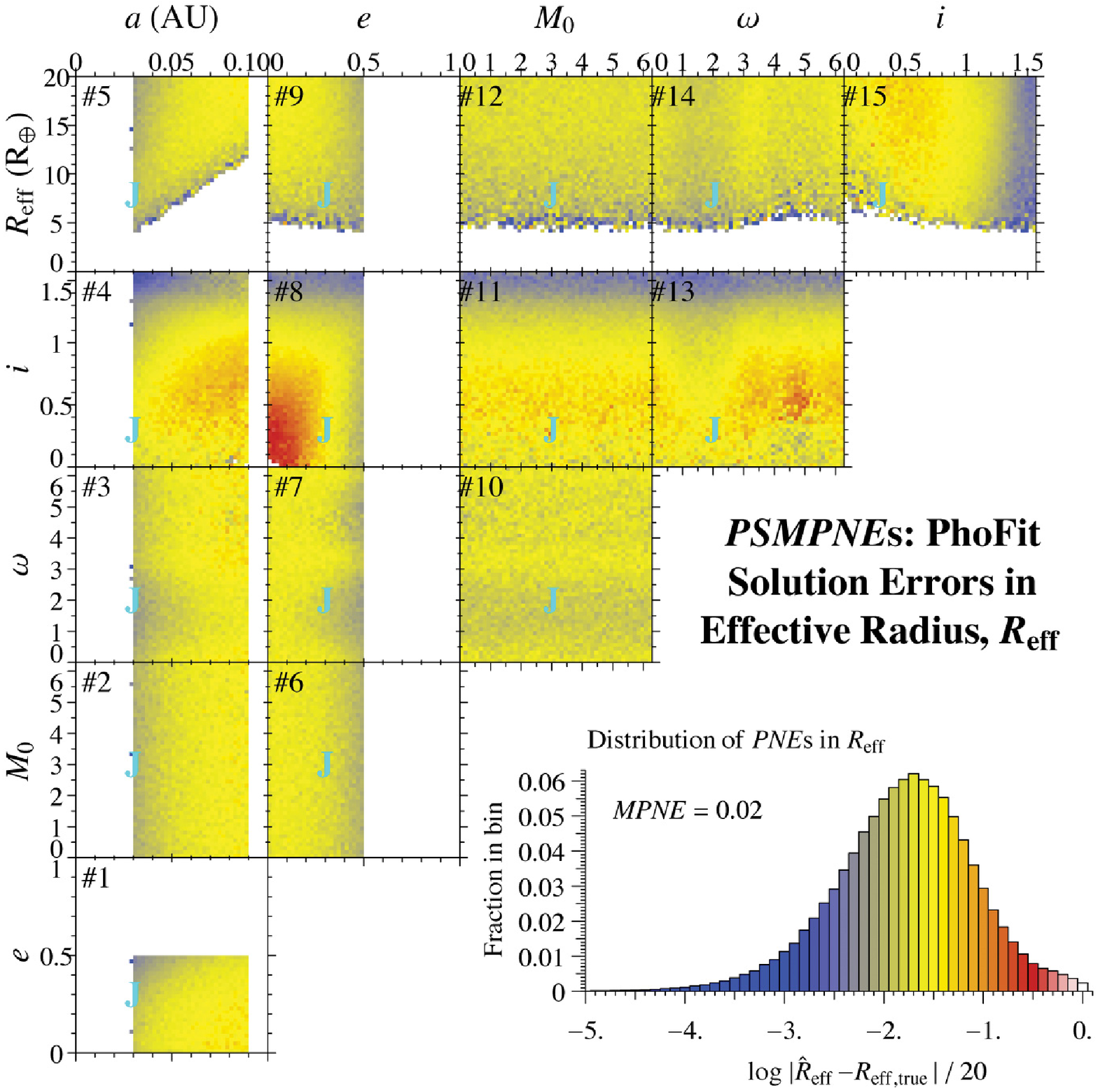}
\caption{\textit{PSMPNEs} for $R_\mathrm{eff}$.}\label{old-f9}
\end{figure}

Figures~\ref{old-f4}--\ref{old-f9} show the paramagrams of the \textit{parameter-specific median projected normalized errors} (\textit{PSMPNEs}):
\begin{equation}
\textit{PSMPNE} \equiv \mathrm{Median} \left[\frac{\textit{fittedValue}-\textit{trueValue}}{\textit{maximumValue}-\textit{minimumValue}}\right]~~,
\end{equation}
where the denominator is the full range of a parameter from Table~1. Table~2 gives the \textit{median projected normalized errors} 
(\textit{MPNEs}) and median absolute errors (MAEs) for all parameters, for all $10^5$ planets considered in this experiment, without regard to location in parameter space. 

Figure~\ref{old-f7}, the paramagrams for the \textit{PSMEs} of $\omega$, consider the universe of well-inclined eccentric orbits ($e$ and $i$  significantly greater than 0). For $\omega\approx\frac{\pi }{2}$ or $\frac{3\pi }{2}$, we are looking at the planet's fullest brightside or fullest darkside, respectively, at periastron, which amplifies light-curve variations that are already most extreme at that time. The distinctiveness of this effect at those geometries reduces the errors in $\omega$, which is signified by the horizontal blue stripes in
frames~3, 7 and 10, and vertical blue stripes in frames~13 and 14. Because the ``backside'' light curve changes more rapidly with changes in $\omega$, the stripes at $\omega\approx\frac{3\pi }{2}$ are bluer (lower error) than those at $\omega\approx\frac{\pi }{2}$.

Consider the \textit{PSMPNEs} for $i$ and $R_\mathrm{eff}$ in Figures~\ref{old-f8} and \ref{old-f9}. We recognize that performing the fits with the observable $h$--$<\!\!h\!\!>$ introduces degeneracy in the form of reciprocity between $R_\mathrm{eff}$ and $i$, which will be more problematic for smaller $e$ and upper--mid-range $i$ and for noisier data sets. In this situation, the value of \textit{peakToTrough}---and approximately the whole light curve---can be preserved by increasing $i$ and appropriately decreasing $R_\mathrm{eff}$, or vice versa. This correlation means $i$  and $R_\mathrm{eff}$ are individually less well constrained---i.e., will show higher errors---in this regime (see both frames~8). When $i$ is smaller, the effect of its variations is larger, and it is better constrained (shows smaller errors); the reverse is true for errors in $R_\mathrm{eff}$, which must fully account for the peak brightness as $i\longrightarrow\frac{\pi }{2}$ (compare the two frames~11). For less noisy data and/or larger $e$, the information carried in the \textit{shape} of the light curve---i.e., the departures from sinusoidality due to \mbox{$\Phi[\beta]$} and/or $e$---helps reduce the uncertainties due to this degeneracy. 

\begin{deluxetable}{lccc}
\tablewidth{0pt}
\tablenum{2}
\tablecaption{\textit{MPNE} and \textit{MAE} are the \textit{median projected normalized}\break and \textit{absolute errors}}
\tablehead{\colhead{Parameter} &\colhead{MPNE} &\colhead{MAE} &\colhead{Units}}
\startdata
Semimajor axis ($a$) &$4\times10^{-6}$ &$2\times10^{-7}$ &AU\\
Eccentricity ($e$) &0.01\phn &0.005\\
Initial mean anomaly ($M_0$) &0.003 &0.02\phn &radians\\
Argument of periastron ($\omega$) &0.005 &0.03\phn &radians\\
Inclination angle ($i$) &0.08\phn &0.1\phn\phn &radians\\
Effective planetary radius ($R_\mathrm{eff}$) &0.02\phn &0.4\phn\phn &$R_{\oplus}$\\
\enddata
\end{deluxetable}

\section{Jupiter Body-Twin}

To examine one PhO in depth, and in order to establish a benchmark planet for the new methods and to demonstrate the basic steps for analyzing \emph{real} data, we used the procedure in \S3 to create a simulated data set for a Jupiter body-twin, in an eccentric, inclined, 1.9-day orbit.  The PhO parameters for this data set (``true solution'') were 
\begin{eqnarray}
P_\mathrm{true} &=&\{a,\,e,\,M_0,\,\omega,\,i,\,R_\mathrm{eff}\}\\ \nonumber
&=&\{0.0300~\mathrm{AU}, 0.300, 3.00~\mathrm{rad}, 5.14159~\mathrm{rad}, 0.300~\mathrm{rad}, 7.90~R_{\oplus}\}~~, 
\end{eqnarray}
for which \textit{peakToTrough}~$=2.7\times10^{-5}$.  The observational parameters and stellar mass were the same as in \S3.  The data set, with the mean subtracted and folded on the period, is shown is Figure~\ref{old-f10}.

Figure~\ref{old-f3} indicates that the periodogram will detect $P_\mathrm{true}$ with 100\% certainty. Nevertheless, the signal is less than one-third the median value for detected planets in the population of Table~1, and therefore significantly weaker than for any of the planets in the wholesale study of \S7. Still, this data set constrains the PhO parameters.

\begin{figure}
\plotone{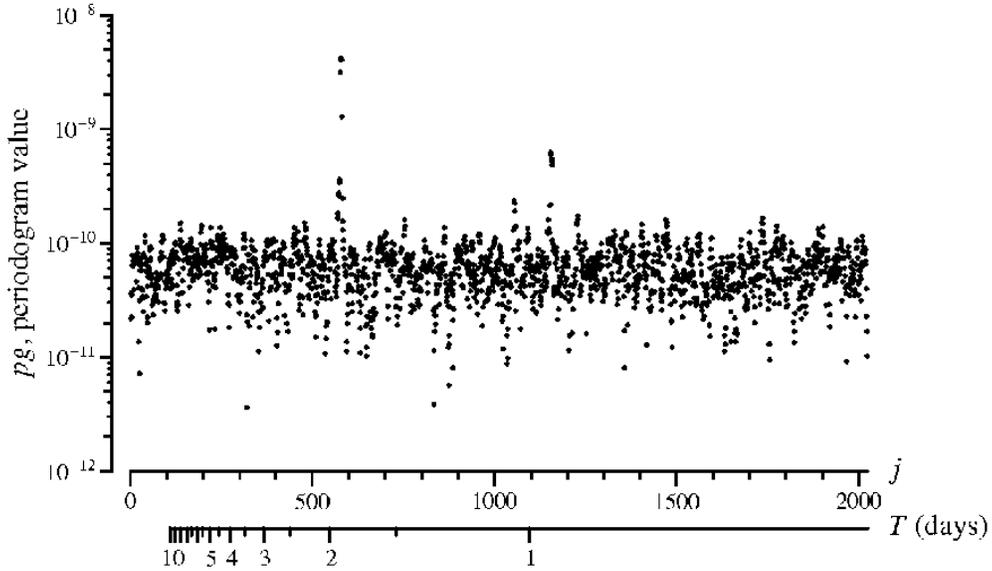}
\caption{Periodogram of the simulated data set created for the Jupiter body-twin. The peak at $T=1.899$~days robustly discovers the planet and estimates the semimajor axis $a=0.03001$~AU with an error of only 0.03\%.} \label{f11}
\end{figure}

Using the protocol of \S4, we find that the periodogram accurately estimates the period at $T=1.899$ days, which corresponds to $a=0.03001$~AU (see Figure~\ref{f11}). 

To find a good starting point to search for a global minimum of \textit{chiSqR}, we evaluated \textit{chiSqR} directly at a large number of  semi-random points in PhO parameter space:  the semimajor axis was held constant at $a=0.0300$~AU, as determined by the periodogram, while $\{M_0,\, \omega,\, i,\, R_\mathrm{eff}\}$ were drawn from uniform random deviates over their full ranges (Table~1), and $e$ was drawn from a uniform random deviate over the range, $0\leq e \leq0.8$.  The lowest value of \textit{chiSqR} discovered was
\newpage
\begin{eqnarray}
\{\textit{chiSqR}_\mathrm{direct,\,min},\ P_\mathrm{direct,\,min}\}&=&\{0.992119, \{0.03000~\mathrm{AU}, 0.300987, 3.04587~\mathrm{rad},\\ \nonumber
&&\qquad 5.13684~\mathrm{rad}, 0.301315~\mathrm{rad}, 7.92696~R_{\oplus}\}\}~~.
\end{eqnarray}

Using $P_\mathrm{direct,\,min}$ as the starting point for the fitting routine, we found
\begin{eqnarray}
\{\textit{chiSqR}_\mathrm{fit,\,min},\ P_\mathrm{fit,\,min}\}&=&\{0.991922, \{0.03000~\mathrm{AU}, 0.28517, 3.11597~\mathrm{rad},\\ \nonumber
&&\qquad 5.14031~\mathrm{rad}, 0.29315~\mathrm{rad}, 8.27574~R_{\oplus}\}\}~~,
\end{eqnarray}
which is identical to the fitting result using $P_\mathrm{true}$ as the starting point.  The value of \textit{chiSqR}$_\mathrm{fit,\,min}$ is consistent with a good fit, a valid theory, and correctly estimated, normal errors.  We would expect to encounter a larger value of \textit{chiSqR}$_\mathrm{fit,\,min}$ some 64\% of the time. (On the same point, the distribution of \textit{chiSqR} for the fits of \S7 closely matched the theoretical reduced $\chi^2$ probability distribution with 4040 degrees of freedom.)

$P_\mathrm{fit,\,min}$ is the best estimate of the PhO. It deviates from the true values by
\begin{equation}
P_\mathrm{fit,\,min} - P_\mathrm{true} = \{2.9\times10^{-7}~\mathrm{AU}, -0.015, 0.116~\mathrm{rad}, -0.00128~\mathrm{rad}, -0.00685~\mathrm{rad}, 0.376~R_{\oplus}\}~~,
\end{equation}
and the fractional deviation in $\sin i$ is 2\%. For real data, $P_\mathrm{true}$ is not known, of course, and we are interested in the confidence regions of the parameters, which we expect will be complex for any marginal Keplerian data sets (Brown 2004b). That is, we expect degeneracies to produce confidence regions with correlations between parameters, and even various \emph{types} of solutions to be compatible with a data set. Such degenerate solutions will evanesce with improving signal-to-noise ratio as more information is gathered.  Each analysis of a real data set is ultimately hand work---teasing out the full information in a data set and finding the optimal manner in which to express and apply it.

\begin{figure}
\plotone{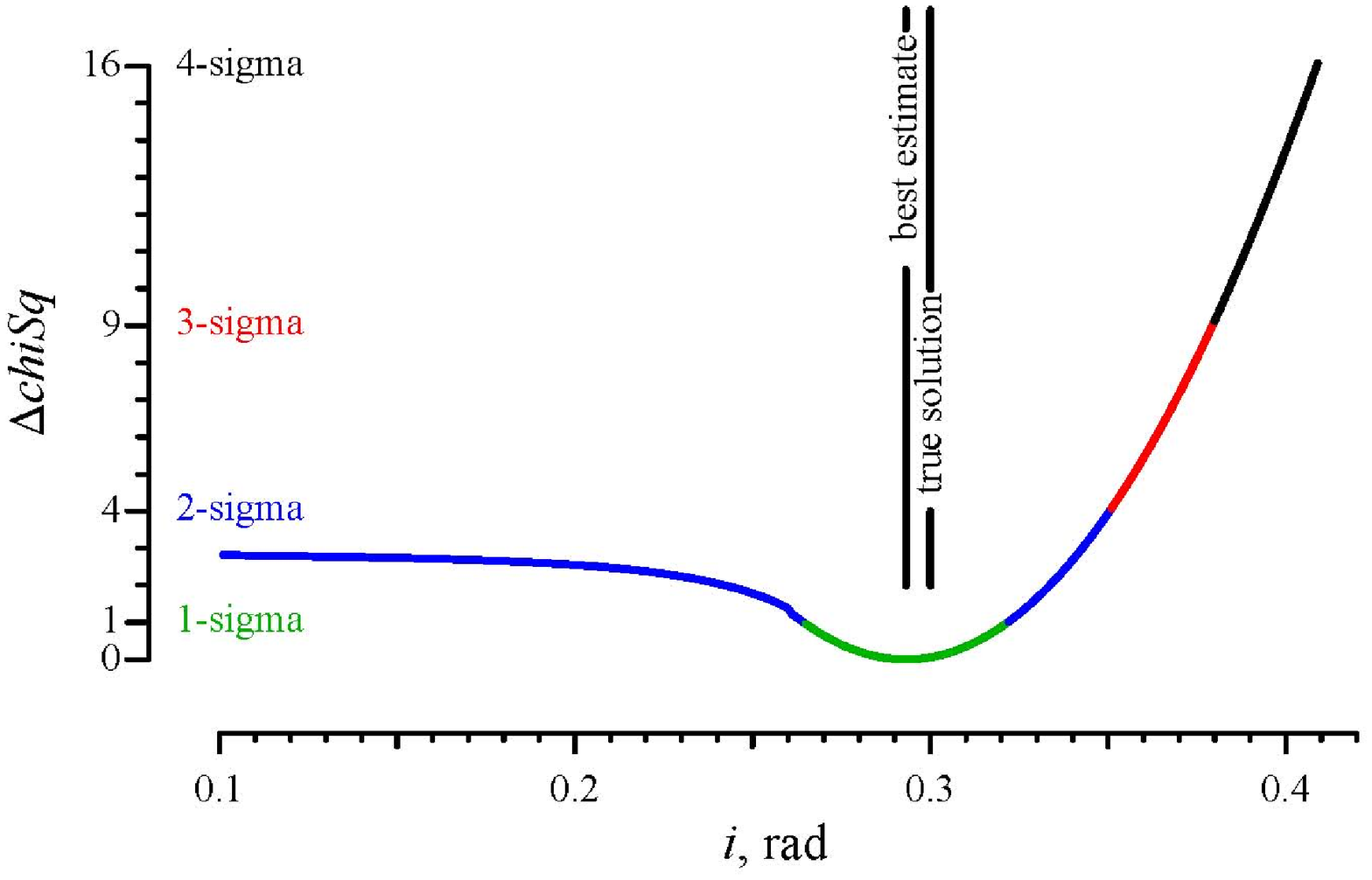}
\caption{Illustration of $\sin i$ recovery for the synthetic data set for a Jupiter body-twin. The confidence region for 1-sigma (68\%) uncertainty in $i$ is less than $\pm$10\%. (Over the indicated range, $\sin i$ differs from $i$ by less than 3\%.) The confidence region becomes asymmetrical at higher confidence levels, demanding a thorough treatment of correlated errors.}\label{f12}
\end{figure}

Because of the special interest in recovering the inclination angle for computing true masses of radial-velocity companions, we computed the confidence regions for $i$ alone, shown in Figure~\ref{f12}.  Here, we have fixed $i$ to values defined by the abscissa, and achieved the best fit to the data by optimizing the other five PhO parameters, then plotted the variation of \emph{unreduced} chi square 
(\textit{chiSq}) from the minimum value
\begin{equation}
\Delta\textit{chiSq}\equiv\textit{chiSq}-4040\,\textit{chiSqR}_\mathrm{fit,\,min}~~.
\end{equation}
$\Delta$\textit{chiSq} is distributed as $\chi^2$ with 1~degree of freedom, which is the same as the probability distribution of the square of a single normally distributed quantity. As shown in Figure~\ref{f12}, the confidence region with $\Delta\textit{chiSq} <n^2$ can be labeled ``$n$-sigma.'' In the current case, with 1-sigma confidence (68\%), we expect the true value of $i$ to lie in the range 0.265--0.320~rad, but with 2-sigma confidence (95\%), we can only say it is less than 0.35~rad. (The results for $\sin i$ are basically the same, because $i$ and $\sin i$ differ by less than 3\% over the range of Figure~\ref{f12}.)

\section{Phase Function Issue}

The handling of the phase function is an issue. All the results in this paper---except in the following paragraphs---used the Lambert phase function in preparing and analyzing simulated data sets. Since the phase function of an SPGP is not known in advance, the question is, what are the effects of analyzing a data set using the wrong phase function?  We explore this question by looking closely at the case of the Jupiter body-twin in \S8.

In Figure~\ref{old-f2}, we introduce a phase-function parameter $\xi$ to control a generalized phase function ranging continuously from lunar ($\xi= -1$) to Lambertian ($\xi= +1$). This range encompasses the variety of phase functions in the Solar System (Sudarsky et~al.\ 2005, Fig.~3). At any given phase angle $\beta$, our generalized phase function interpolates linearly between the values at phase angle $\beta$ of the lunar and Lambert phase functions. 

\begin{figure}
\epsscale{.65}
\plotone{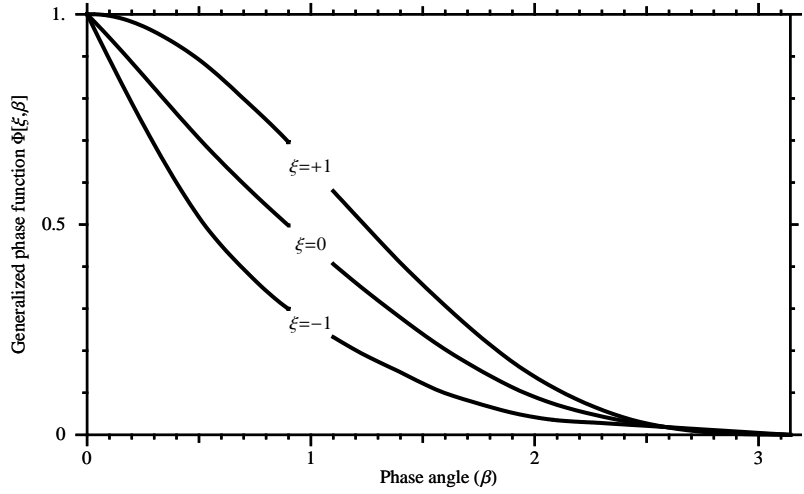}
\caption{The generalized phase function developed for investigating the consequences of using the wrong phase function when estimating orbital parameters. $\xi$ is the phase-function parameter. The lunar phase function 
($\xi = -1$) is taken from Figure~3 in Sudarsky et~al.\ (2005). The Lambert phase function ($\xi = +1$) is given in Eq.~3. For intermediate values of $\xi$, the phase function at any phase angle $\beta$ is a linear interpolation between the lunar and Lambert values at $\beta$.}
\label{f13}
\end{figure}

It is apparent from Figure~\ref{f13} that the estimated value of $R_\mathrm{eff}$ will increase if we use for orbit fitting a phase function with a lower value of $\xi$ than the true value. Lowering $\xi$ reduces theoretical values for $\Phi$, which must be compensated by larger theoretical values of $R_\mathrm{eff}$. Furthermore, Eq.~1 suggests that the increase in $R_\mathrm{eff}$ should be proportional to the change in the square root of the effective phase integral, $I[\xi]$, where
\begin{equation}
I[\xi] \equiv \int^\pi_0 \sin\beta\, \Phi[\xi,\beta]\, Q[\beta]\, d\beta~~,
\end{equation}
where $Q[\beta]$ is the probability distribution of the phase angle ($Q[\beta]$  is different for each orbit). If we have a good PhO fit using an arbitrarily selected phase function $\xi_1$, we could roughly predict the fitted value of $R_\mathrm{eff}$ if we had used any other phase function, $\xi_2$, by multiplying $R_\mathrm{eff}[\xi_1]$ by $\sqrt{I[\xi_1]/I[\xi_2]}$.

\begin{figure}
\epsscale{.65}
\plotone{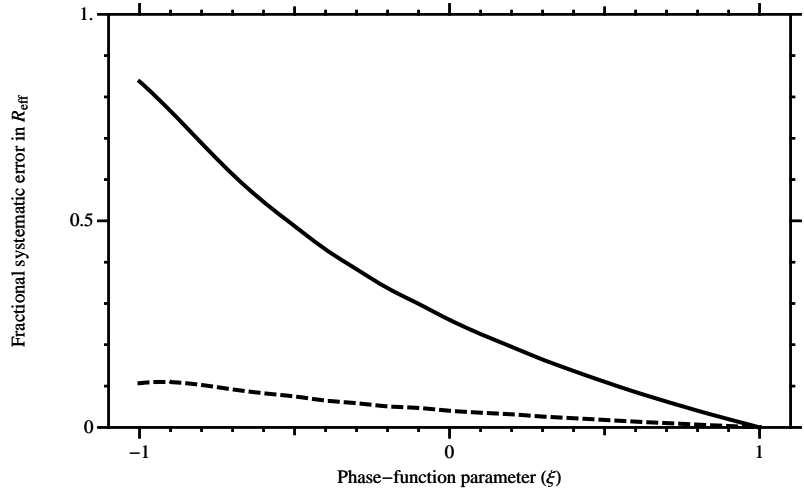}
\caption{Solid line: the fractional systematic error in $R_\mathrm{eff}$ when the Jupiter body-twin data set (\S8), which was prepared using the Lambert phase function, is analyzed with the generalized phase function with parameter $\xi$. Dashed line: remaining fractional error after the estimate of $R_\mathrm{eff}$ is corrected by the square root of the change in the effective phase integral.}\label{f14}
\end{figure}

\begin{figure}
\epsscale{.7}
\plotone{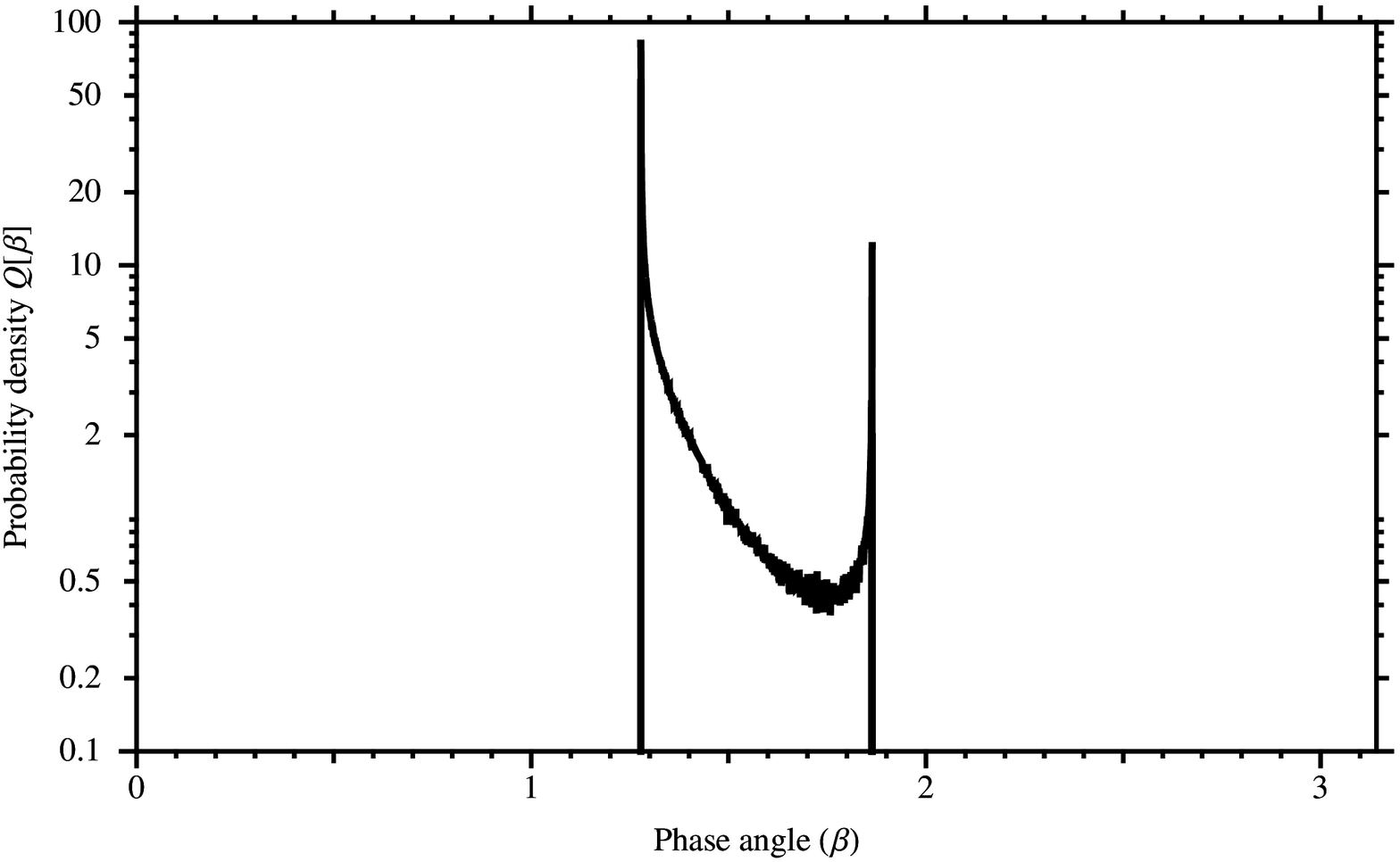}
\caption{Empirical probability distribution of the phase angle $\beta$ for the Jupiter body-twin data set studied in \S8. To obtain this curve, we evaluated $\beta$ at one million random epochs, prepared an empirical cumulative probability distribution, took the numerical derivative, and verified the normalization to unity of the resulting $Q[\beta]$. We used $Q[\beta]$ to evaluate the effective phase integral for the Jupiter body-twin according to Eq.~27. }\label{f15}
\end{figure}

The solid curve in Figure~\ref{f14} shows the fractional systematic error in $R_\mathrm{eff}[\xi]$ when the Jupiter body-twin data set is analyzed using phase functions in the range $-1 \le \xi \le +1.$ Figure~\ref{f15} shows $Q[\beta]$ for the Jupiter body-twin orbit. The dashed curve in Eq.~14 shows the fractional 
deviation of $R_\mathrm{eff}[\xi]\sqrt{I[\xi_1]/I[\xi_2]}$ from $R_\mathrm{eff} [\xi_1]$. This correction indicates a basic understanding of the dominant factor in the phase-function issue in regards to $R_\mathrm{eff}$. 

\begin{figure}
\epsscale{.7}
\plotone{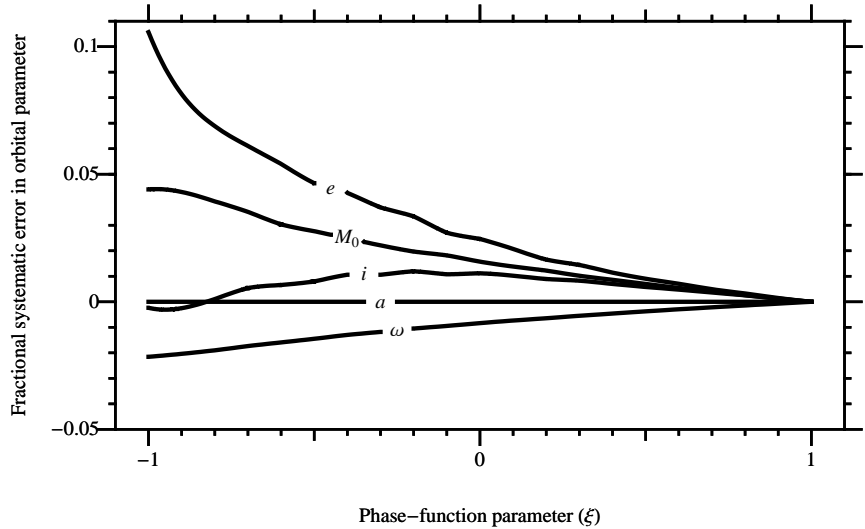}
\caption{Systematic errors in the PhO parameters $a$, $e$, $M_0$, $\omega$, and $i$ when the Jupiter body-twin data set is analyzed using a phase function different than the Lambert phase function ($\xi = 1$) with which the data set was prepared. In the most important case of $i$, the systematic errors are smaller than the typical random errors for similar PhOs (see Fig.~\ref{old-f8}).}\label{f16}
\end{figure}

Figure~\ref{f16} shows the systematic errors due to the phase function for the other five orbital parameters in the case of the Jupiter body-twin data set. The semimajor axis $a$ is not affected, because it is controlled by the periodicity of the signal. The systematic errors in $e$, $M_0$, and $\omega$ are somewhat larger than the typical random errors of similar PhOs (see Figs.~\ref{old-f5}--\ref{old-f7}. For the inclination angle, $i$, which is perhaps the most important PhO parameter, the systematic error is \emph{smaller} than the typical random errors: $<$2\% compared with typically $\sim$10\% (see Fig.~\ref{old-f8}).

For a real data set, a practical approach would be simply to fit the data using the best guess as to phase function, and to qualify the result with that choice. Because the phase function itself can only be estimated from data with much higher $S/N$ than afforded by \textit{Kepler}, we will need to be content for now with estimates of $e$, $M_0$, $\xi$, and $R_\mathrm{eff}$ that depend on the assumed phase function. The most complete treatment will specify an \textit{a~priori} probability distribution for the phase function (i.e., for the phase-function parameter $\xi$), in which case the combined error for each orbital parameter could be expressed as the convolution of the probability distributions of systematic and random errors.

Because of the somewhat smaller systematic error in inclination angle $i$ as compared to its random errors, our prediction that \textit{Kepler}-level photometry may break the $m \sin i$ degeneracy in a useful manner is not substantially undermined by the phase-function issue. 

\section{Discussion and Conclusion}
We have defined and put forward an approach to investigating the PhO, which we believe will come to be recognized as a Keplerian entity on a par with radial-velocity and astrometric orbits.

The PhO solves in principle one of the outstanding problems in astronomy: the $m \sin i$ degeneracy in radial-velocity observations. The way is now open to computing true masses for radial-velocity companions, starting with SPGPs. 

Brown (2009) shows for large, noisy Keplerian data sets---particularly in the case of astrometry, but radial velocity and photometry are all basically the same in this regard---that search completeness via periodogram, as well as the completeness of accurate estimates of orbital parameters by least squares, are solely functions of the signal-to-noise ratio ($S/N$). $S/N$ is defined as the semi-amplitude of the signal times the square root of the number of data points, divided by the single-measurement error. For example, 50\% search completeness calls for $S/N=5$--8, due to any combination of signal amplitude, number of data points, and single-measurement error (so long as the orbit is well covered by the measurements). (The color scale above the abscissa in Figure~4 shows $S/N\approx7$ for the PhO.) High completeness for, say, $\pm10$\% accuracy in estimates of orbital parameters typically calls for $S/N=50$--60. In the case of the Jupiter body-twin data set shown in Figure~\ref{old-f10} and treated in \S8, $S/N=48$. Therefore, detection is guaranteed for this planet for the assumed \emph{Kepler} performance and observing parameters. Furthermore, the high $S/N$ of the Jupiter body-twin data set means that the small deviations of the fitted orbital parameters from the true values (Eq.~25) are no surprise.

PhO observations may be difficult. There is uncertainty about how low the 
albedos of SPGPs actually are, about their phase functions, and about a possible 
noise floor due to systematic errors from instrumental and stellar sources. 
Nevertheless, simple detection of SPGPs in reflected light should be robust in 
the regime of \emph{Kepler} photometry, and estimates of all six orbital parameters may be feasible in at least a subset of cases.

Burrows et~al.\ (2008) offers the most recent discussion of observed and theoretical values of geometric albedo $p$ for the case of SPGPs. The best measured value, upper limit $p = 0.038 \pm 0.045$ of Rowe et~al.\ (2008), from \textit{MOST} observations of HD~209458b, is compatible with cloudless models. The same models suggest an that $p$ could increase by factor two at the somewhat longer wavelengths sampled by \textit{Kepler}. Meanwhile, transit photometry of SPGPs has revealed astonishingly large radii---1.74 $R_\mathrm{jupiter}$ in the case of TrES-4 (Mandushev et~al.\  2007). A planet with the size of TrES-4 and twice the geometric albedo of HD~209458b would offer $R_\mathrm{eff} = 5.4 \pm 5.8 R_\mathrm{earth}$. If that planet were on the orbit of the Jupiter body-twin, the signal-to-noise ratio would be $S/N=22^{+74}_{-22}$.

As expected for any new method, particularly one debuted with such epochal claims as breaking the $m \sin i$ degeneracy, our treatment of the PhO will certainly attract questions and concerns, which we must study and address. We would be the first to caution about systematic effects, particularly about a noise floor that might limit the averaging out of random measurement errors. Nevertheless, the potential payoff of determining the inclination angles of radial-velocity companions independently---that alone---is so great that we must press forward.   Whatever its current technical limitations, we can expect the grasp of this new method to improve with time, as instrumentation improves and systematic errors are addressed.

A key issue is how well our method stands up to more realistic noise, particularly due to stellar and planetary variability. The latter must wait for \emph{Kepler} results and the possibility that observations may discover variations in planetary properties. The former, however, is a known issue, which we must study using more realistic random deviates for the photometry. That research will likely produce selection criteria for the future target lists.  We expect that a meaningful subset of \emph{Kepler} targets will support PhO modeling. This expectation is supported by the prediction of Jenkins \& Doyle (2003) that \emph{Kepler} will discover ``from 100 to $\sim$760 SPGPs'' in reflected light by periodogram searching. All these SPGPs are potential candidates for PhO analysis.

It has been asked how our simple model and six-parameter solutions relate to the apparent complexities of the \textit{forward problem},
that is, predicting photometric properties from detailed planetary models (Seager et~al.\ 2000; Dyudina et~al.\ 2005; Sudarsky et~al.\  2005; Gaidos et~al.\ 2006). We are observers, working on the \textit{reverse problem}, which is inferring planetary properties from a photometric record. Unless the signal-to-noise ratio of \emph{Kepler} data sets is much higher than we expect, we predict satisfactory $\chi ^2$ probabilities for fits using only six parameters, which would say that additional parameters are not required by the observations.  For systematic effects with assumed values that are not constrained by the observations, the associated range of uncertainty must be convolved with random errors when the total uncertainty in estimates are expressed, as we have outlined in \S9 the case of the phase function.


\begin{acknowledgements}
We thank Adam Burrows for his comments on the manuscript. We thank Christian Lallo for aiding the computations and assisting with the graphics. We thank Sharon Toolan for her expertise preparing the manuscript. JPL contract 1289452 with the Space Telescope Science Institute provided support for this research.
\end{acknowledgements}

\end{document}